# Dayside and nightside magnetic field responses at 780 km altitude to dayside reconnection

K. Snekvik[1], N. Østgaard[1], P. Tenfjord[1], J. P. Reistad[1], K. M. Laundal[1,2], S. E. Milan[1,3], and S. E. Haaland[1,4]

[1]Birkeland Centre for Space Science, University of Bergen, Bergen, Norway, [2]Teknova AS, Kristiansand, Norway, [3]Department of Physics and Astronomy, University of Leicester, Leicester, UK, [4]Max Planck Institute for Solar System Research, Göttingen, Germany

**Abstract** During southward interplanetary magnetic field, dayside reconnection will drive the Dungey cycle in the magnetosphere, which is manifested as a two-cell convection pattern in the ionosphere. We address the response of the ionospheric convection to changes in the dayside reconnection rate by examining magnetic field perturbations at 780 km altitude. The Active Magnetosphere and Planetary Electrodynamics Response Experiment data products derived from the Iridium constellation provide global maps of the magnetic field perturbations. Cluster data just upstream of the Earth's bow shock have been used to estimate the dayside reconnection rate. By using a statistical model where the magnetic field can respond on several time scales, we confirm previous reports of an almost immediate response both near noon and near midnight combined with a 10–20 min reconfiguration time of the two-cell convection pattern. The response of the ionospheric convection has been associated with the expansion of the polar cap boundary in the Cowley-Lockwood paradigm. In the original formulation of this paradigm the expansion spreads from noon to midnight in 15–20 min. However, also an immediate global response has been shown to be consistent with the paradigm when the previous dayside reconnection history is considered. In this paper we present a new explanation for how the immediate response can be accommodated in the Cowley-Lockwood paradigm. The new explanation is based on how MHD waves propagate in the magnetospheric lobes when newly reconnected open flux tubes are added to the lobes, and the magnetopause flaring angle increases.

## 1. Introduction

The two-cell ionospheric convection pattern during southward interplanetary magnetic field (IMF) is well established [e.g., *Heppner and Maynard*, 1987; *Weimer*, 1995; *Haaland et al.*, 2007]. It was predicted by *Dungey* [1961] as a consequence of the large-scale magnetospheric convection. It has also been known for a long time that the convection responds within 10 min to a southward turning of the IMF [*Nishida*, 1968]. A framework for explaining the time-dependent response of the ionospheric convection changes is given by the paradigm of *Cowley and Lockwood* [1992] (hereafter referred to as CL). See also *Cowley and Lockwood* [1997] and *Lockwood and Morley* [2004].

In the CL paradigm the convection commences in two cells a few minutes after the onset of dayside reconnection at the magnetopause. Due to the ionospheric incompressibility, the ionospheric convection streamlines have to be closed on timescales of about a few seconds [*Vasyliūnas*, 2012]. After about 1 h, the convection is further enhanced by reconnection in the magnetotail during the substorm expansion phase [e.g., *Bargatze et al.*, 1985]. The two cells form near the ionospheric footprint of the dayside X line, and in the next 10–20 min they gradually expand toward midnight. A corresponding time delay should therefore be observed between the convection responses on the dayside and the nightside. Such delays had been reported several times in the years leading up to the CL paradigm [*Lockwood et al.*, 1986; *Etemadi et al.*, 1988; *Todd et al.*, 1988; *Saunders et al.*, 1992]. Note that the reconfiguration time is much shorter than the convection time across the polar cap which is typically 1–2 h during steady southward IMF [*Zhang et al.*, 2015].

Later studies, on the other hand, found time delays which appeared to disagree with the CL paradigm [*Ruohoniemi et al.*, 2002]. *Ridley et al.* [1998] found no time delay in the ground magnetic field responses and proposed that the two-cell convection pattern reaches its final form in less than a minute. Afterward, the pattern is fixed, but the convection continues to increase in magnitude for about 13 min on average. Also, other studies have reported results which can be interpreted differently than the CL paradigm suggests. *Jayachandran and MacDougall* [2000], *Murr and Hughes* [2001], *Lu et al.* [2002], and *Nishitani et al.* [2002] found the initial responses to be simultaneous at all local times they observed, but the times of maximum responses were more delayed at the nightside. This can be interpreted as being due to an evolving convection pattern. *Murr and Hughes* [2001] and *Lu et al.* [2002] both pointed out that it is important to separate between the times of initial and maximum responses to correctly describe the reconfiguration of the convection pattern. In this paper we will use the term rise time to quantify the difference between the maximum and initial response times. The rise time near midnight is typically 10–20 min longer compared to near noon. More recently, however, *Fiori et al.* [2012] found that also the initial convection responses are delayed near midnight. *Anderson et al.* [2014] found a response in field-aligned currents at the nightside 20–30 min delayed compared to the dayside. *Morley and Lockwood* [2006] have analyzed the evolving convection pattern during time-varying magnetic reconnection and found that a global response can often be consistent with the CL paradigm. Interested readers are recommended the comprehensive overview of this topic given in their introduction.

Ambiguous observations of the response time have come from the magnetosphere as well. *Wing et al.* [2002] investigated the response of magnetic fields at the geosynchronous altitude and identified a time delay of 7 min between the initial responses at the dayside and nightside magnetosphere. On the other hand, *Nishimura et al.* [2009] found no systematic time delay between the initial convection response in the inner magnetosphere at the dayside and the nightside, in agreement with the simultaneous response of the partial ring current at different local times found by *Hashimoto et al.* [2002].

In this paper we make use of 10 min accumulated global maps with 2 min cadence of magnetic field perturbations $\mathbf{B}_\perp$ above the high-latitude ionospheres. These measurements are used to determine the time development of the ionospheric convection velocity $\mathbf{V}$ at nine different local times between noon and midnight. $\mathbf{B}_\perp$ are associated with the magnetic stress which sustains plasma convection in the collisional ionosphere [e.g., *Iijima*, 2000]. In an equilibrium situation, without any neutral wind, the stresses would be perfectly balanced by the frictional force:

$$\frac{1}{\mu_0} B_0 \frac{\partial \mathbf{B}_\perp}{\partial Z} \pm \rho \nu_{in} \mathbf{V} = 0 \,, \tag{1}$$

where $B_0$ is the geomagnetic field, $Z$ is the vertical direction, $\rho$ is the plasma mass density, and $\nu_{in}$ is the ion neutral collision frequency. It is assumed that the magnetic field perturbation $\mathbf{B}_\perp$ is small compared to $B_0$, and that the geomagnetic field is approximately radial. The sign depends on the hemisphere. Thus, one might expect $\mathbf{B}_\perp$ and $\mathbf{V}$ to be parallel and antiparallel above the southern and northern ionosphere, respectively. *Tu et al.* [2014] showed that this is not necessarily the case because of the Hall effect, but the results presented below strongly indicate that $\mathbf{B}_\perp$ is nearly aligned with $\mathbf{V}$ on average.

As noted by *Ruohoniemi et al.* [2002], a challenge with using observations to determine the response times in single events is the large sample-to-sample variability of the ionospheric convection. This is particularly true for the Active Magnetosphere and Planetary Electrodynamics Response Experiment (AMPERE) data set due to the low resolution of the Iridium magnetometers. This uncertainty—and the large number of event studies which have concluded differently about the response time—have motivated us to use a statistical approach to determine the relevant timescales of the responses relative to changes in the dayside reconnection rate. These timescales include the time of the initial response, the time delays of the succeeding enhancements of the response, and the exponential rise time of the response at each local time. This is achieved by using an autoregressive model with exogenous inputs, with the solar wind reconnection rate as the exogenous input, and $\mathbf{B}_\perp$ as the output.

The details of the method, as well as the data selection criteria are described in the next section. An average map of $\mathbf{B}_\perp$ is presented in section 3, and it is discussed how the convection response can be determined. A case study illustrating the use of the technique is shown in section 4, and section 5 contains the statistical results. An interpretation of the results in terms of MHD waves is given in section 6, and we suggest an alternative to *Morley and Lockwood* [2006] for how there can be a global onset of ionospheric convection. In section 7, we discuss our results, and in section 8 we give the main conclusions.

## 2. Data and Methods

### 2.1. Data

For this study it is crucial to use solar wind data as close to the Earth as possible. For this purpose we have used Cluster data just upstream of the bow shock. Solar wind intervals have been identified from ion data [*Rème et al.*, 2001] and magnetic field data [*Balogh et al.*, 2001] from Cluster 4. These data sets are used to calculate the effective dayside reconnection rate [*Kan and Lee*, 1979; *Milan et al.*, 2008]:

$$R = |V_X B_{YZ}| \sin^2 \frac{\theta}{2}, \quad \tan \theta = \frac{|B_Y|}{B_Z}, \quad 0 \le \theta < \pi \qquad (2)$$

In this case, the reconnection rate is the magnetic flux which reconnects per X line unit length per unit time [Wb/(m s)].

The AMPERE data set [*Waters et al.*, 2001; *Anderson et al.*, 2014] provides magnetic field perturbations and corresponding field-aligned currents from a low orbit of 780 km. The data set is based on a spherical harmonic fit to measurements from the Iridium Communications constellation. Each fit is based on all measurements in a 10 min data window. The measurements are obtained from 66 satellites distributed in six orbit planes which each carries a low-resolution engineering magnetometers with a resolution of 48 nT. The spherical fit provides a spatial resolution of 3° in latitude and 2.4 h in longitude. Each 10 min data window is separated by 2 min.

The time tags are centered in each sampling interval for both data sets. The solar wind data has been time shifted for 8 min to take into account the solar wind propagation through the magnetosheath from the bow shock nose to the magnetopause [*Ridley et al.*, 1998; *Slinker et al.*, 2001].

### 2.2. Selection Criteria

Each event must fulfill the following criteria:

1. The event should be within 1 month from equinox. This makes the conditions in the Northern and Southern Hemispheres as equal as possible, which simplifies the comparison between the hemispheres.
2. Cluster should be continuously in the solar wind for at least 12 h. Fifty-seven such time intervals were identified in 2011 and 2012.
3. The reconnection rate was required to exceed 2 mWb/(m s) at least once per time interval. Many substorms occur during such driving [*Pulkkinen et al.*, 2007]. This criterion reduced the number of time intervals to 30.

### 2.3. Regression Model

The purpose of this paper is to determine how the ionospheric convection responds and reconfigures after a change in the flux transport from the dayside reconnection region. Since the magnetosphere-ionosphere system can respond on multiple timescales, a model is needed which separates between the response times and the rise times of the response. For these purposes we have used an autoregressive model of first order with the dayside reconnection rate as exogenous input, which can be written as

$$B(n) = cB(n-1) + \lambda_1 R(n-l_1) + \lambda_2 R(n-l_2) + \cdots + \lambda_m R(n-l_m) + e(n). \qquad (3)$$

$B$ and $R$ are the magnetic field perturbation above the ionosphere and the dayside reconnection rate, respectively. The first term on the right-hand side is the autoregressive term. The constant $c$ represents the rise time of the response as will be shown below, and $n$ refers to one measurement in one of the 12 h intervals. The times of the initial response and the subsequent changes of the response are given by the lags as $2l_i$ min since the sampling interval is 2 min. We will refer to these times as the response times. The lags $l_i$ can take values of 0–45, which means that the modeled magnetic field is predicted from 0 to 90 min earlier values of $R$. In the statistical analysis presented in section 5 it was found that the number of significant lags $m$ can be between 2 and 4 in the different local time sectors. The coefficient $\lambda_i$ quantifies the change of the response for the different lags. A positive value means an enhancement, while a negative value means a decrease of the response. The residuals or errors, denoted by $e(n)$, is the difference between the measured and modeled magnetic field for sample number $n$. This is a special case of the more general class of NARMAX models [*Billings*, 2013].

The parameters $c$, $\lambda_i$, and $l_i$ are determined by the method of orthogonal least squares with forward regression [*Chen et al.*, 1989]. The method can be summarized as follows.

1. In forward regression one input parameter is determined at a time. The first step is to determine the model of the form of equation (3) with only one lag of $R$:

$$B(n) = cB(n-1) + \lambda_1 R(n-l_1) + e(n). \qquad (4)$$

Forty-six different values of the sum of squared errors $\sum e(n)^2$ are calculated by letting the lag $l_1$ vary between 0 and 45. The lag $l_1$ which gives the smallest sum of squared errors is kept for the next step.

2. The next step is to determine the second lag $l_2$ which, in combination with the first lag $l_1$, minimizes the errors:

$$B(n) = cB(n-1) + \lambda_1 R(n-l_1) + \lambda_2 R(n-l_2) + e(n). \tag{5}$$

Again, $\sum e(n)^2$ is calculated for all possible values of the lag $l_2$, and the lag which gives the smallest sum of squared errors is kept for the next step.

This procedure can be repeated until the number of desired lags terms are found. Note that the coefficients $c$ and $\lambda_i$ are recalculated at each step. In sections 4 and 5 we will discuss how many lags are significant.

The method described above can be extended to multiple time intervals by the method "Multiple orthogonal search (MOS) for model term selection" by *Wei and Billings* [2009]. The goal is to find common lags for all the time intervals, which minimize the errors. The coefficients are then calculated as the mean of the coefficients from all the time intervals found in section 2.2:

$$\lambda_i = \frac{1}{J} \sum_{j=1}^{J} \lambda_{i,j}, \tag{6}$$

where $J$ is the number of time intervals.

### 2.3.1. Rise Time ($t_r$)

Equation (3) without any error term, and with constant input, is a first-order difference equation. To see this, let the input be written as $\beta = \sum_i \lambda_i R(n-l_i)$. Equation (3) becomes $B(n) = cB(n-1) + \beta = c^n B_0 + \beta(1-c^n)/(1-c)$, where $B_0 = B(n=0)$. This equation will increase or decrease asymptotically to

$$B(\infty) = \beta/(1-c) \text{ if } c < 1. \tag{7}$$

The relative difference between $B(n)$ and $B_0$ can be written

$$\frac{B(n) - B_0}{B(\infty) - B_0} = 1 - c^n = 1 - \exp\left(-\frac{t}{t_r}\right), \tag{8}$$

where $t_r$ is the rise time defined as the time it takes for the difference to reach $1 - e^{-1} = 63\%$ of the asymptotic difference for constant input. This gives a rise time of $t_r = -2/\ln(c)$, since $t = 2n$.

## 3. Average Map of $\mathbf{B}_\perp$

Figure 1 shows the average maps of FACs and magnetic field perturbations in the Northern and Southern Hemispheres in Altitude Adjusted Corrected Geomagnetic (AACGM) coordinates *Baker and Wing* [1989]. The arrows show $\mathbf{B}_\perp$. The colors correspond to field-aligned currents (FAC) with red away from the Earth and blue toward the Earth. The Southern Hemisphere is seen through the Earth. The averages have been calculated for the 30 intervals with Cluster in the solar wind. The components of the IMF are distributed as follows: $B_X = 0.6 \pm 3.7$ nT, $B_Y = -1.4 \pm 4.8$ nT, and $B_Z = -1.0 \pm 4.1$ nT in GSM. The second values are the standard deviations.

The region 1 and 2 currents are clearly seen in Figure 1, likewise the sunward and antisunward directions of the magnetic field perturbations. The direction of the arrows corresponds well to the two-cell convection pattern during southward IMF. In the north the arrows are antiparallel with the expected flow direction since the geomagnetic field is pointing downward, in accordance with equation (1).

The average $\mathbf{B}_\perp$ is larger between the region 1 and 2 currents than in the polar cap. This simply reflects the higher conductivity and plasma density there, giving a larger friction force and a larger magnetic field perturbation, as seen from equation (1). We will use the magnetic perturbations equatorward of the polar cap to estimate the convection response to dayside magnetopause reconnection in different magnetic local time (MLT) sectors. Only arrows with azimuthal components corresponding to sunward convection are used to estimate the responses.

The sectors where the responses are calculated are marked in Figure 2. The sectors contain the local times 02–10 MLT on the dawnside and 14–22 MLT on the duskside. The sectors lie between 60° magnetic latitude and the poleward edge of the region 1 current. The average location of the region 1 current is shifted toward midnight and slightly toward dusk as seen from Figure 1. The poleward edge can be approximated by a circle with 13° radius and with the center shifted 2.2° toward midnight and 0.9° toward dusk. Average values are calculated based on pairs of MLT sectors at opposite sides of the midnight meridian. For example, one average is calculated for the 10 MLT and 14 MLT sectors combined, another average is calculated for the 09 MLT and 15 MLT sectors combined, and so on.

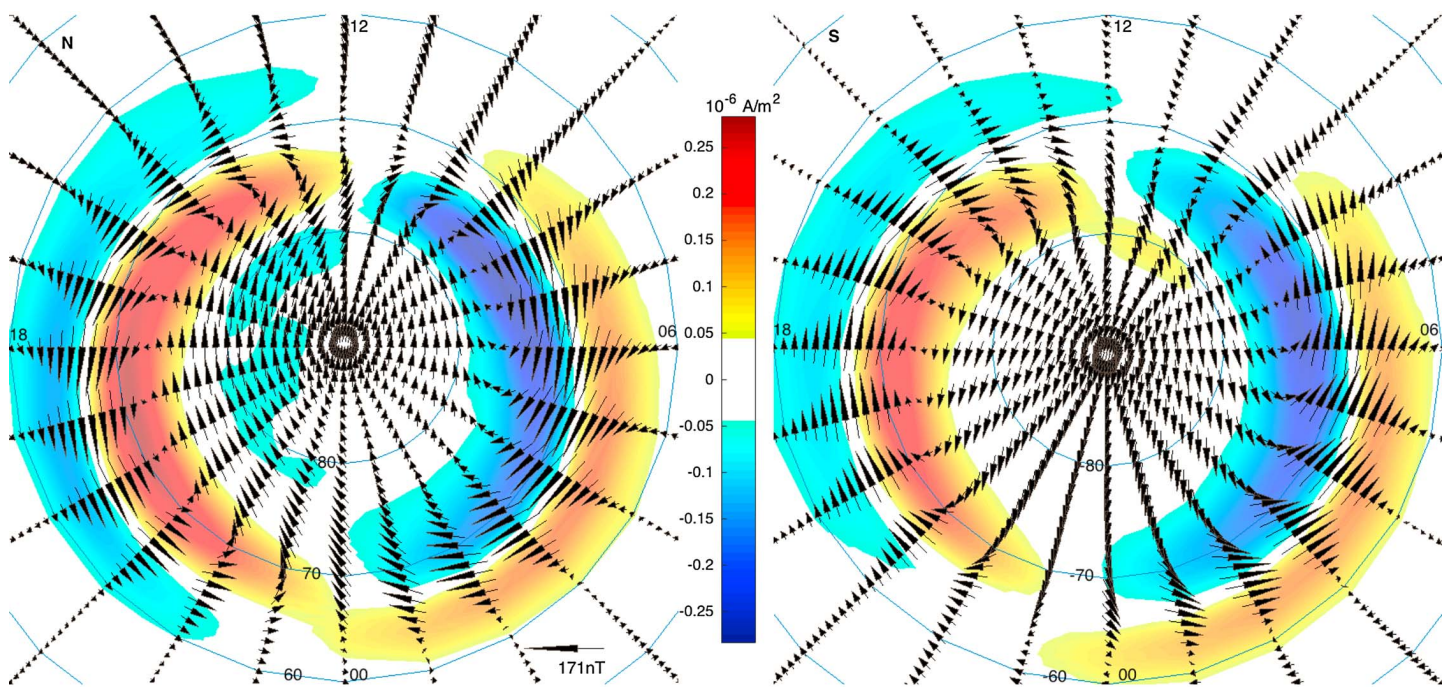

**Figure 1.** Average maps of FACs and magnetic field perturbations from the (left) Northern and (right) Southern Hemispheres.

This approach would not work for other convection cells than the two-cell convection pattern. In section 7 we discuss the effect of the lobe cell and the viscous cell and argue that they will not significantly alter the results which will be reported below.

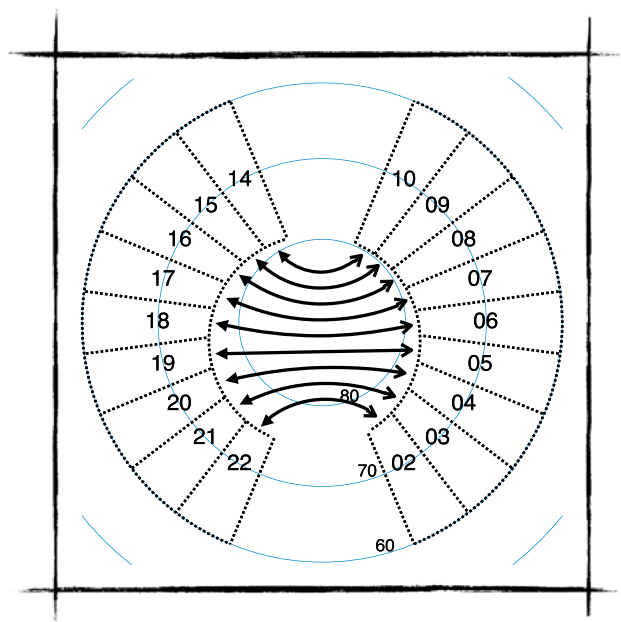

**Figure 2.** The MLT sectors where the magnetic responses have been calculated. Each sector is combined with the sector on the opposite side of the midnight meridian as indicated by the arrows. The number inside each sector corresponds to MLT.

## 4. Case Study: 1 March 2011

Figure 3 shows the magnetic responses from one of the 12 h time intervals with Cluster 4 in the solar wind. The solar wind velocity along GSM $X$ is shown in Figure 3a. All three GSM components of the IMF are shown in Figure 3b. The dayside reconnection rate (equation (3)) is shown in Figure 3c. The magnetic perturbations $B$ in five MLT sectors from noon to midnight are shown in Figurez 3d–3h with red lines. The blue lines are outputs from the regression model and will be explained later. Two MLT sectors at each side of the midnight meridian are combined as explained in the previous section.

During this event the IMF turns quickly southward two times and stays negative for more than one hour. The first southward turning is just after 05:00 UT, and the second southward turning is just before 09:00 UT. During both turnings *R* increases sharply. As such, the event should be well suited to study the magnetic response with AMPERE and the time delays between the dayside and nightside responses.

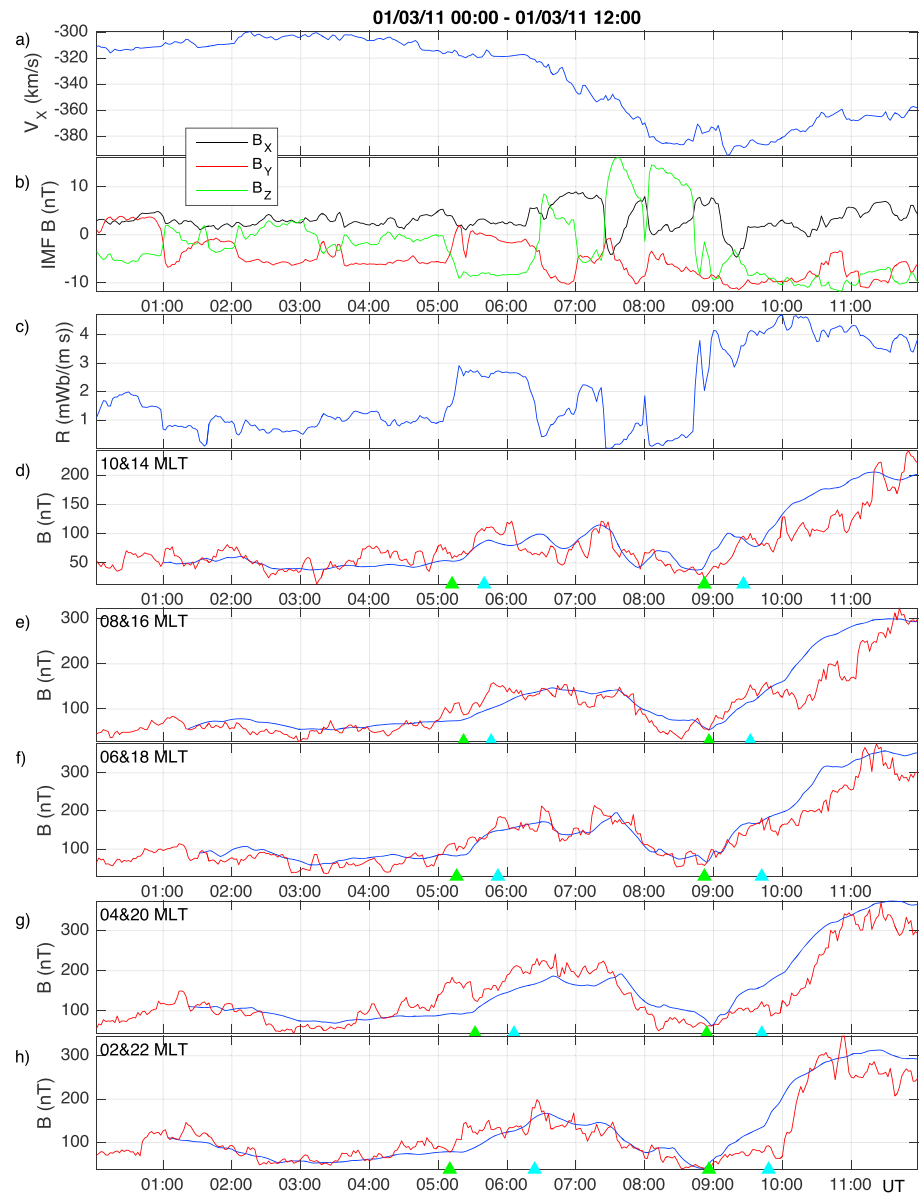

**Figure 3.** The three top panels show: (a) The solar wind velocity along GSM *X*, (b) all three GSM components of the IMF, and (c) the magnetopause reconnection rate field *R*. (d–h) The red lines show the magnetic responses from AMPERE corresponding to sunward flow in five MLT sectors in the Northern Hemisphere. The blue lines are the output from the regression model. Green and cyan triangles suggest initial and full response times to the southward turnings.

Before the southward turning at 05:10 UT, IMF $B_Z$ fluctuates around zero for several hours. However, a strongly negative IMF $B_Y$ contributes to an *R* around 1 mWb/(m s). Despite the stable IMF, the AMPERE magnetic field fluctuates with an amplitude of about 50 nT, which is comparable to the resolution of the magnetometers wherein the measurements were obtained from. This makes it difficult to determine exactly when the magnetic response starts. As an approximation, we characterize the responses by the initial and full response times. The initial response time in each sector is the first enhancement after the southward turning. And the full response time is the subsequent maximum of *B*. In the five MLT sectors the initial responses seem to be at 05:12 UT, 05:22 UT, 05:16 UT, 05:32 UT, and 05:10 UT when sorted from noon to midnight. The maxima seem to be at 05:40 UT, 05:46 UT, 05:52 UT, 06:06 UT, and 06:24 UT. Thus, only the maximum responses are more delayed toward midnight. The times for the initial and full response times are shown with colored triangles.

Before the southward turning at 08:44 UT, the IMF has been northward for more than 70 min except for a brief southward excursion at 08:00 UT. For this turning it is easier to identify the initial responses in the different MLT sectors. In the five MLT sectors they seem to be at 08:52 UT, 08:56 UT, 08:52 UT, 08:54 UT, and 08:56 UT when sorted from noon to midnight. The initial responses in the two sectors closest to midnight are quite weak, but they both increase strongly after about an hour. These delayed responses seem to be associated with a substorm starting at 10:30 UT with $AL$ reaching below −500 nT (not shown). There are local maxima in all MLT sectors about 40 min after the initial response. They seem to be at 09:26 UT, 09:32 UT, 09:42 UT, 09:42 UT, and 09:48 UT. Again, only the time delay of the maximum response in each sector shows a systematic increase toward midnight.

## 4.1. Regression Results for the Case Study

This case study illustrates how difficult it is to precisely determine the response times of the magnetic field to the reconnection rate when there is large variability in the data. Furthermore, we will show in section 5.1 that it is only possible to accurately determine the initial response times statistically due to the low resolution of the magnetometers. Identifying the times is necessarily subjective. It is therefore desirable to have a completely objective way to estimate the response times. This is the motivation for the regression model of this study.

For this case study we chose to fit a regression model with three lags of $R$ in equation (3). In the statistical analysis in the next section we will determine how many lags are significant. All the estimated parameters for the regression models are given in Appendix A. The response times $2l_i$ min and the corresponding coefficients $\lambda_i$ are illustrated in Figure 4. Negative coefficients are shown by crosses, and positive coefficients are shown by circles. The relation between the size of the symbols and the magnitude of the coefficients are shown in the legend in the top of the figure. The outputs of the regression models are also shown with blue lines in Figure 3d-3h. The initial response times seem to agree better after the second southward turning, compared with that after the first southward turning. However, the data increase slower than the regression output near midnight after 09:00 UT. We have no good explanation for this.

Negative coefficients ($\lambda_i < 0$) will act to reduce the response, while positive coefficients ($\lambda_i > 0$) will act to increase the response. As seen from equation (3) it is the weighted sum of the past reconnection rates $R$ which determines whether the response increases or decreases. Two main timescales of positive coefficients stand out in Figure 4. The first is after 6–14 min, and the second is after 56–88 min. The response time of the initial timescale increases from 6 min in the 10&14 MLT sector to 14 min in the 03&21 MLT sector but decreases to 8 min closest to midnight.

The rise times of the responses are 15–37 min, with 28 min as the average value (see the Appendix A). The effect of the long rise times and short response times is that the magnetic field responds quickly to a change in the reconnection rate, but takes a long time to completely reconfigure. We will discuss the effect of the long rise time later.

It can be shown that for each additional time lag which is added to equation (3), the improvement to the model is smaller than it was for the previously added time lag [*Billings*, 2013]. Therefore, the order in which the time lags occur are shown with colors in Figure 4. The primary, secondary, and tertiary lags are shown with black, brown, and pink colors, respectively. The primary lags all belong to the first timescale, while all the positive secondary lags belong to the second timescale.

In the 05&19 MLT sector and the 02&22 MLT sector there are additional positive coefficients delayed with about 10 min after the initial response. They both belong to the tertiary lags. Due to the low resolution of the Iridium magnetometers it is challenging to discriminate between so short response times. In addition, four of the coefficients in Figure 4 are negative. This leads to the important question of how many input terms to use in equation (3). If the goal is to get an optimal good fit, one should use as many terms as possible, but this would lead to a lot of terms which are just fitted to the sample-to-sample variability of the data. However, all the input terms should ideally represent physical processes connected to the magnetosphere-ionosphere convection. The statistical analysis in the next section is intended to get rid of the "noise" terms and to find the mean response times corresponding to the "process" terms.

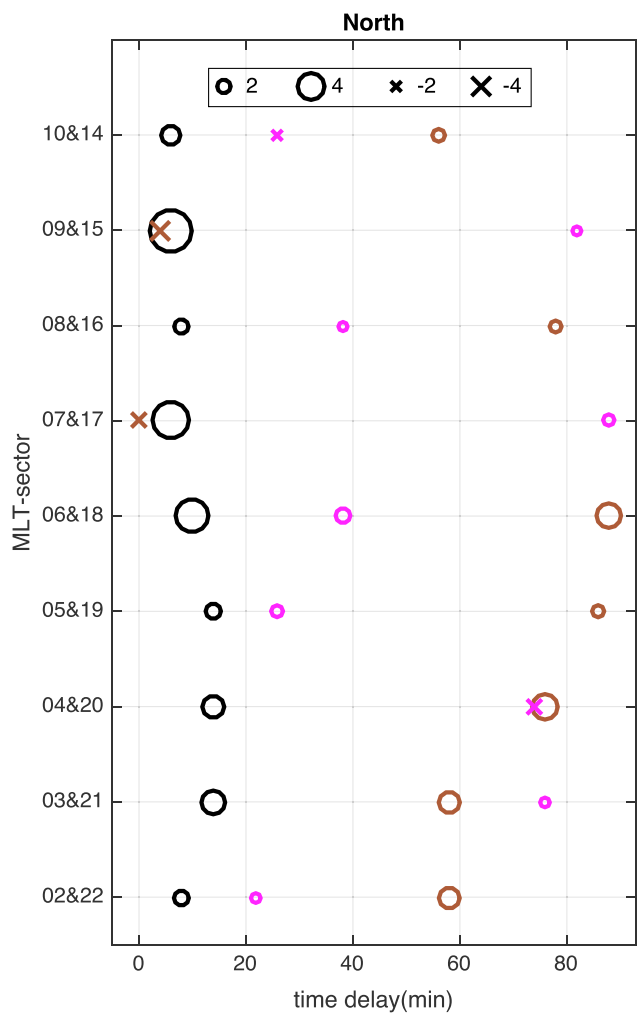

**Figure 4.** Visualization of the response times $2l_i$ and the corresponding coefficients $\lambda_i$ from equation (3) for the case study. The magnitudes of $\lambda_i$, with units nT/(mWb/(m s)), are shown in the legend at the top of the figure, where negative and positive coefficients are shown by crosses and circles, respectively. Different MLT sectors are on the vertical axis, and the response times are on the horizontal axis. The colors correspond to the order the lags $l_i$ were found in the forward regression: (1) black, (2) brown, and (3) pink.

## 5. Statistical Results

The first step in the statistical analysis was to identify and exclude events with atypical magnetic field response to the reconnection rate. Three time lags of $R$ in equation (3) were used as in the case study. These terms were determined for the 30 twelve hour intervals with Cluster in the solar wind by the method described in section 2. The coefficients $c$, $\lambda_1$, $\lambda_2$, and $\lambda_3$ were determined separately for each time interval, resulting in 30 values for each coefficient for each of the nine MLT sectors in each hemisphere. The mean values of the coefficients in each MLT sector were calculated from equation (6), with $J = 30$. The deviation of each set of coefficients from the mean values was estimated based on the Mahalanobis distance [e.g., *Storch and Zwiers*, 1999, chap. 2]. If it is assumed that the coefficients are normally distributed, the Mahalanobis distance has a $\chi^2$ distribution with $4°$ of freedom. We defined a set of coefficients to be atypical if it was outside a 99% confidence interval from the mean values, in at least three different MLT sectors.

All the atypical events were removed, and the procedure described above was repeated until no more atypical events were left. In total three atypical events were found. Closer inspection of the solar wind data revealed plausible explanations for removing these events. In the first event the proton density increased by more than 400%, and this lasted for several hours. In the second event IMF $B_Z$ was strongly northward in the last hours of the event. In the third event IMF $B_Z$ was negative throughout the event, and the reconnection rate was not varying very much.

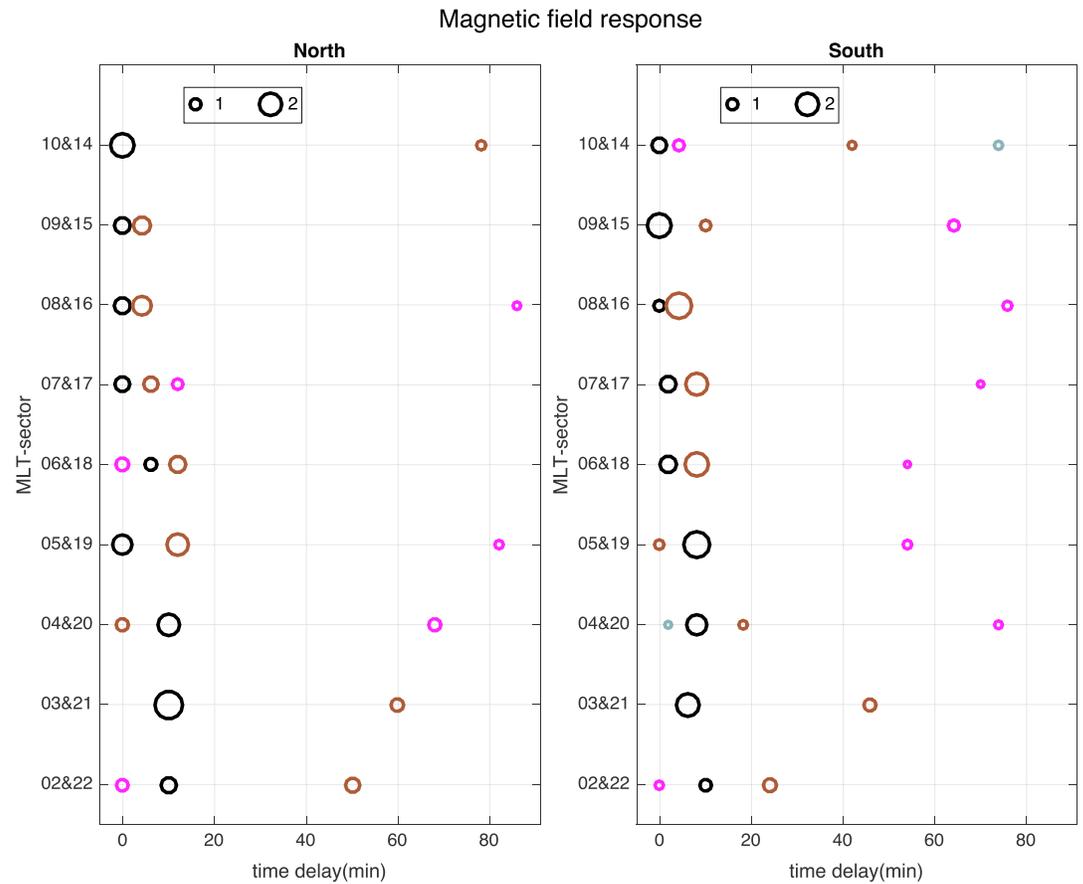

**Figure 5.** Visualization of the response times $2l_i$ and the corresponding coefficients $\lambda_i$ for the statistical study in the same format as Figure 4. The order of the lags are given as (1) black, (2) brown, (3) pink, and (4) grey. No negative coefficients were found in the statistical study.

After all the atypical events had been removed, the time lags and the corresponding coefficients were estimated again. For each coefficient a 95% confidence interval for the mean value was estimated based on the Student's $t$ statistic. The maximum number of time lags with coefficients significantly different from zero were determined for each MLT sector. Note that this way of identifying the model size differs from the method described in *Wei and Billings* [2009]. The results are shown in Tables A2–A3 in Appendix A.

Just as in the case study, there is one group of rapid response times and another group of more delayed response times, as can be seen in Figure 5. The first group of response times is after 0–12 min and the second group is after 42–86 min. Only two response times fall outside these two groups. Unlike the case study, there are no negative coefficients ($\lambda_i$). This indicates that the negative coefficients were due to the sample-to-sample variability in the data.

There appears to be two subgroups among the rapid response times. The first subgroup consists of very short response times of 0–2 min. We will refer to this as the instantaneous response times. This is seen in all, but one of the MLT sectors in both hemispheres. The second subgroup consists of response times delayed by 8–12 min. Only 4 of 17 and 3 of 17 responses fall outside these two categories in north and south, respectively. Thus, the response times agree with those studies who have found a global rapid onset of convection combined with a reconfiguration of the convection pattern during 10–20 minutes [e.g., *Lu et al.*, 2002]. Due to the 10 min data windows of AMPERE, we can only consider response times separated by 10 min or more to be truly independent. In the following we will therefore focus on the response times with zero delay and the response times with 10 min delay or more.

The order in which the lags occur in the forward regression is shown by colors in Figure 5, just as in Figure 4. The primary lags correspond to the instantaneous response times in the four MLT sectors closest to noon in both hemispheres. But in the three sectors closest to midnight, the primary lags correspond to the shortly delayed response times. The coefficients of the shortest time lags are also decreasing in magnitude away from noon with approximately 0.09 nT/(mWb/(m s)) per hour in the north and 0.12 nT/(mWb/(m s)) per hour in the south.

The rise times of the responses varied between 34 and 49 min in the north and between 28 and 45 min in the south. The long rise times have the effect that it can be difficult to distinguish between the instantaneous and the shortly delayed response times. This is discussed in the next section.

Some idea of the accuracy of the response and rise times can be found by comparing the northern and southern polar cap. The response times at 0–8 min and 10–18 min agree very well in the north and south. The response times above 40 min have more variation between the north and south, which implies an uncertainty of tens of minutes. The difference between the rise times in north and south is about 10 min or 20% on average.

### 5.1. Model Magnetic Field

Equation (3) minus the error term gives the model magnetic field. Figure 6 shows the modeled magnetic field for selected MLT sectors from the Northern Hemisphere for an artificial created solar wind input. The dayside reconnection rate ($R$, Figure 6a) is set to zero from the start. After 2 h it increases as a step function to 3 mWb/(m s) and remains at this value to the end.

This figure clearly shows the effects of the short response times and the long rise times. While the outputs respond very quickly to the increased $R$, the curves are still increasing substantially after more than 1 h. The figure illustrates that it can be very difficult to distinguish between two response times which follow each other closely in time due to the long rise time. For example, in the sector closest to midnight (02&22 MLT), Figure 5 shows that there are three response times at $2l_1 = 10$, $2l_2 = 50$, and $2l_3 = 0$ min. The effect of the delayed response times is a steepening of the slope of the curve in Figure 6f. This would be hard to identify if there was additional variability in the magnetic field data.

Closer inspection of Figure 6f reveals that the increase of the magnetic field is only 14 nT between the two first response times. This is less than one third of the 48 nT which can be resolved by the magnetometers on board the Iridium satellites. It might seem controversial that the regression analysis gives results which are superior to the resolution of the magnetometers. However, it is the power of statistics which makes this possible, and we will illustrate this with an example in Figure 7.

The reconnection rate in Figure 7a is the same as in Figure 6a, except that the increase is set to 2 mWb/(m s), which is more common in the solar wind. Figure 7b shows the modeled magnetic field response in the 02&22 MLT-sector, and Figure 7c shows what a magnetometer with 48 nT resolution would measure. This shows that it would not be possible to distinguish between the different response times from the magnetometer data in one single event.

Figure 7d shows 40 different profiles of the reconnection rate with initial values of 0–1 mWb/(m s). Each profile increases with 2 mWb/(m s) after 2 h as in Figure 7a. The modeled responses are shown in Figure 7e, and Figure 7f shows the corresponding hypothetical measurements by a magnetometer with 48 nT resolution. Finally, the average of the "measurements" are shown in Figure 7g. The response times of 0, 10, and 50 min from Figure 6f and 7b are clearly reproduced in the average profile. This shows that the resolution of the magnetometer is not an absolute limitation when considering the statistics of many events as we have used in this paper.

## 6. Interpretation

There are three main results from this paper which need to be explained in a self-consistent way: (1) The magnetic field responds simultaneously in all MLT sectors but with decreasing magnitude away from noon. (2) The magnetic field on the nightside is further enhanced after 10 min. (3) The magnetic field increases gradually with a rise time of 28–49 min. In addition, there is a third response time of 42–86 min which agrees very well with the observed duration of the substorm growth phase [e.g., *Iyemori*, 1980]. Since our purpose is to study the reconfiguration of the ionospheric convection pattern as a direct response to dayside reconnection, we will not discuss this response time any further. Our results will be interpreted in terms of the ionospheric convection. The uncertainty due to the ionospheric conductivity is discussed in section 7.

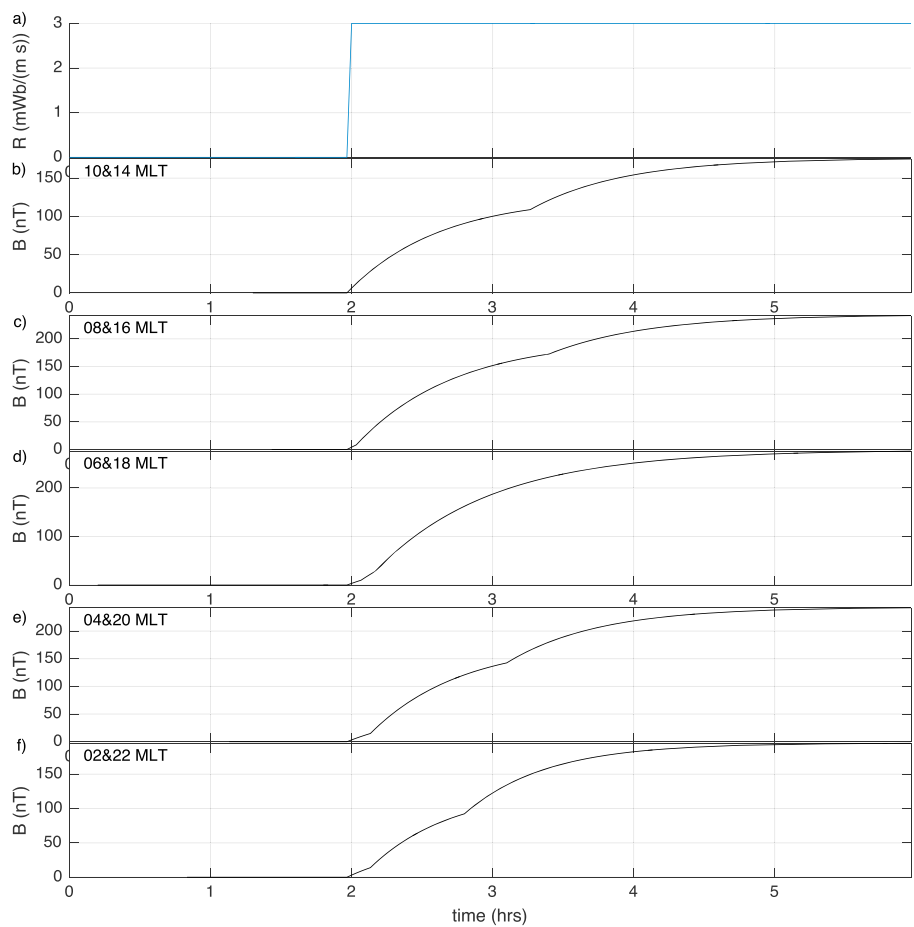

**Figure 6.** Magnetic field from the regression model for artificial solar wind input for selected MLT-sectors in the Northern Hemisphere.

During dayside reconnection, magnetic flux is added to the magnetospheric lobes. As long as the dayside reconnection rate is higher than the nightside rate, the magnetic flux in the lobes will increase [*Cowley and Lockwood*, 1992]. To accommodate the larger flux, the magnetotail radius must increase, giving a larger magnetopause flaring angle [*Petrinec and Russell*, 1996]. This will increase the perpendicular force on the magnetopause from the solar wind bulk pressure, and the lobe magnetic field increases due to pressure balance across the magnetopause [*Caan et al.*, 1975]. It is well known that a pressure perturbation in a plasma propagates as a compressive wave. The pressure gradient at the compression front will accelerate the plasma perpendicular to the magnetic field and thus initiate the convection in the magnetospheric lobes. A similar argument can be made for the rarefaction wave in the closed magnetosphere [*Slinker et al.*, 2001], but we will focus on the open field lines for the sake of brevity.

*Tamao* [1964] has derived the equations for the hydromagnetic waves with a localized source in a cold plasma and discussed how the waves will propagate from a source region on the magnetopause. If the source is a point source, the disturbance will spread in all directions and the disturbance will decrease with the distance squared [*Tamao*, 1964, equation 4.3]. It was also shown that the compressive disturbance from a localized source will be a source of shear Alfvén waves propagating parallel to the field lines. The shear Alfvén wave will transfer the momentum between the magnetosphere and the ionosphere. Note that for plane waves, there is no coupling between the compressive waves and the shear Alfvén waves. The coupling found by *Tamao* [1964] is due to the localized source.

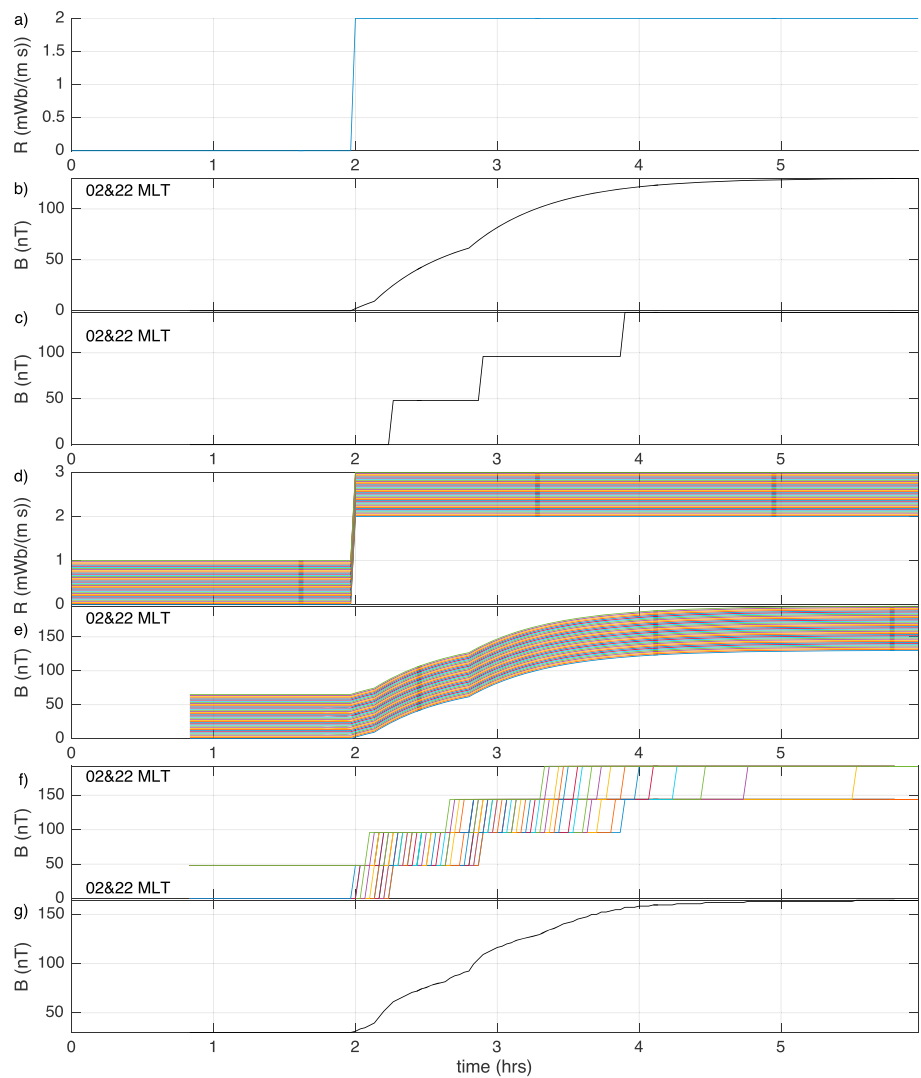

**Figure 7.** (a) The reconnection rate. (b) The magnetic field response from the regression model. (c) The measurements by a hypothetical magnetometer with 48 nT resolution. (d–f) Same as above for 40 different profiles of the reconnection rate. (g) The average of the "measurements" in Figure 7f.

Based on these concepts we will interpret the two first response times as illustrated in Figure 8. The times $t = 0$ and $t = 10$ min are relative to that when reconnection has started on the dayside. At $t = 0$ a small flux tube has been swept tailward from the $X$ line by the solar wind and added to the lobes. This increases the flaring angle very locally at the magnetopause. The new positions of the magnetopause are shown by dotted lines in Figures 8a and 8b. As explained above, a compressive wave (yellow lines) will propagate away from the magnetopause. At $t = 0$ the source is close to a point source, and the amplitude of the compression decreases rapidly with distance from the source. The compressive wave will excite plasma flow (red arrows) not only toward the plasma sheet but also toward the flanks of the lobe and thus gradually increase the flaring angle over a larger area on the magnetopause. After a while, the source of the compressive wave is no longer localized to a point. This is illustrated for $t = 10$ min. At this time the plasma flow should be less attenuated with distance from the magnetopause.

The lobe convection will be transmitted by shear Alfvén waves to the ionosphere. This is illustrated by the blue field aligned arrows in Figure 8 b). At $t = 0$ the amplitudes of the Alfvén waves are larger near noon than near midnight, while at $t = 10$ min the amplitudes are much more similar. Magnetic tension in the ionosphere will accelerate the plasma, and this disturbance propagates to surrounding field lines as a compressive wave. The disturbance spreads out in the entire ionosphere in a few seconds [*Song and Vasyliūnas*, 2014] and creates the two-cell convection patterns.

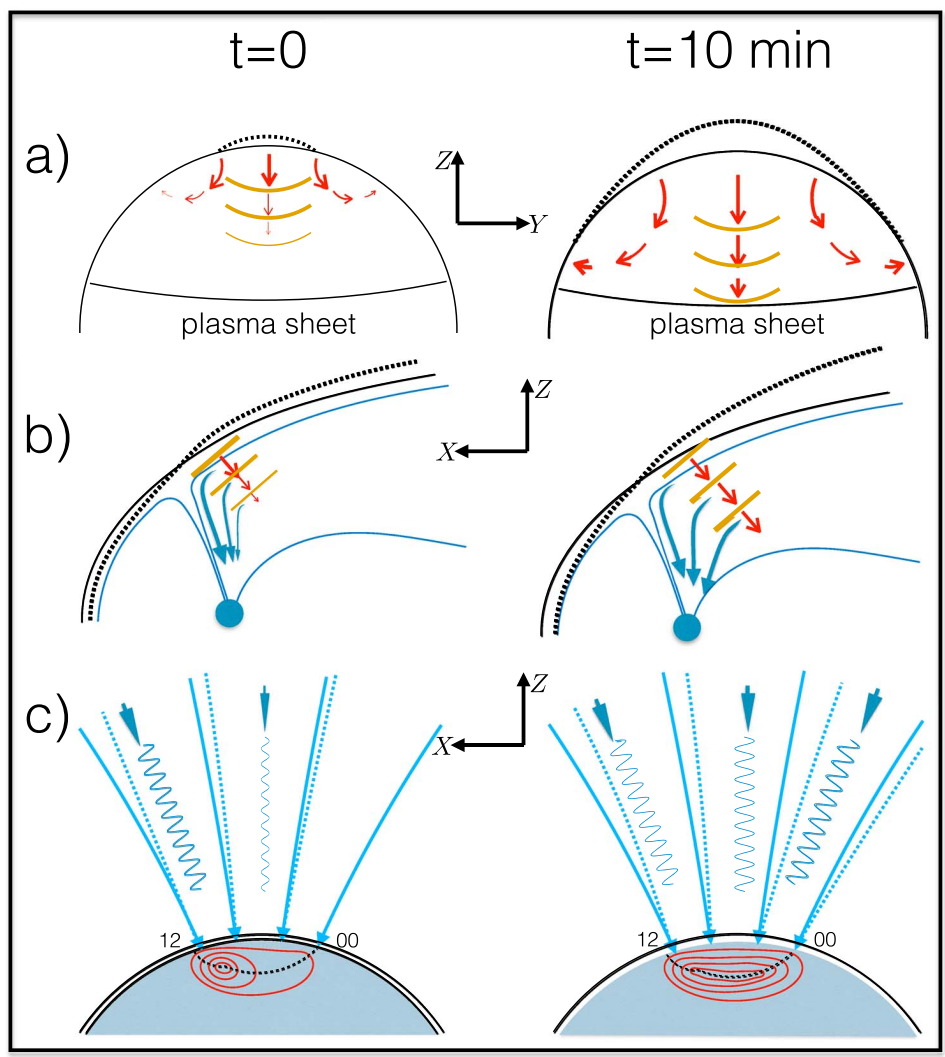

**Figure 8.** Reconfiguration of the ionospheric convection pattern the first 10 min after a southward turning of the IMF. (row a) A cross section of the near-Earth magnetotail as seen from the tail. (row b) A cross section of the magnetosphere in the noon-midnight meridional plane. The solid and dotted lines represent the magnetopause position before and after reconnection. The red arrows are the convection, and the yellow lines are phase fronts of the fast mode. The Alvén mode is represented by the blue field-aligned arrow. (row c) The situation near the high-latitude ionosphere in the same plane as Figure 8 (row b). The blue solid and dotted lines show the unperturbed field lines and the perturbed field lines, respectively.

Two dusk convection cells consistent with this interpretation and with the response times at 0 and 10 min are drawn in Figure 8c). The magnetic stress is such that the plasma convects into the polar cap near the cusp in accordance with the expanding-contracting paradigm of *Cowley and Lockwood* [1992]. The polar cap boundary will be shifted equatorward by the new open flux, and the surrounding closed field lines will flow back toward the reconnection region. At $t = 0$ the incident shear Alfvén waves are stronger near noon compared to near midnight, and the resulting convection cell is centered postnoon. Only a few stream lines extend to midnight consistent with the weak response there. At $t = 10$ the incident Alfvén waves are more equally distributed in the polar cap, and the resulting convection cell is symmetric across the dawn-dusk meridian.

In the discussion above, the time delays due to the propagation of the waves have been ignored. The relevant delays are the propagation times through the lobe of the compressive wave and along the field lines of the shear Alfvén wave. Both waves propagate with the Alfvén speed in a cold plasma. Assuming that the propagation times of the Alfvén waves are approximately the same near noon and near midnight, it is only the propagation time through the lobe which will make a difference. Due to an Alfvén speed of a few thousand km/s, this delay is approximately 1 min, which is an order of magnitude less than the 10 min response time we want to explain.

The rise time of the magnetic response has often been interpreted as the number of times an shear Alfvén wave will reflect in the ionosphere before the ionospheric and magnetospheric convection becomes equal [*Goertz et al.*, 1993; *Tu et al.*, 2014]. However, we will argue that at least part of the rise time can be explained by the interaction between the solar wind and the magnetosphere. The gradual increase of the magnetopause flaring angle provides such a mechanism. The magnetopause flaring angle will continue to increase as long as more flux is added to the lobes by dayside reconnection than what is removed by nightside reconnection.

To further understand the meaning of the rise time, it is instructive to write equation (3) as the change in the magnetic field during one time step.

$$\Delta B(n) = (c-1)B(n-1) + \sum_i \lambda_i R(n-l_i) + e(n). \tag{9}$$

The first term on the right is always negative, while the second term is always positive. If we continue to interpret the magnetic field in terms of the convection, it is clear that the first term can be interpreted as a decelerating force, while the second term is the accelerating force described above.

Consider a situation similar to the one discussed in section 5.1 and shown in Figure 6. Just after dayside reconnection has started, the first term is small, and $\Delta B$ is large. After a while when $B$ has increased, the first term becomes of similar magnitude as the second term, and the convection will be closer to a steady state. From this argument it follows that a high value of $c$ corresponds to a small decelerating force and a long rise time.

It is important to note here that the regression model was only chosen due to its ability to estimate response times and rise times separately. So we cannot expect that there is any real physical force which corresponds to the first term. Still, we have identified two mechanisms which could give a decelerating force which increases with time. The first mechanism could be the friction force in the ionosphere. As seen from equation (1) the friction force is proportional to the convection speed. It is the friction force which contributes to the Joule heating in the ionosphere [*Strangeway and Raeder*, 2001], and the Joule heating is the major energy sink in the magnetosphere-ionosphere system [*Tenfjord and Østgaard*, 2013]. The other decelerating force could be an effective reflection of the compressive wave near the plasma sheet. This happens because the lobe field lines cannot effectively pass the plasma sheet barrier [e.g., *Dmitrieva et al.*, 2004]. When field lines start to accumulate near the lobe-plasma interface, another compressive wave would propagate away from the plasma sheet effectively reducing the acceleration of the lobe plasma.

## 7. Discussion

### 7.1. Model

The regression model used here has some advantages as well as some limitations. Many previous publications have used cross correlation to infer the response times in the ionosphere. However, *Lu et al.* [2002] have pointed out that cross correlation will not always give correct time delays between noon and midnight when the rise time depends on local time. Only the initial response times can be used to infer the expansion of the convection cells. Different methods to find these times have been discussed by *Morley and Lockwood* [2005]. One limiting factor for this is that the signal has to rise above the background noise level before it can be identified. Another limiting factor in the case of AMPERE is that the resolution of the magnetometers is only 48 nT. Our analysis indicates that it takes about 30 min for the magnetic field perturbations to increase to this level for a reconnection rate of 2 mWb/(m s). One should be careful with using AMPERE to determine response times during shorter intervals except in statistical studies.

In our model the response times and the rise time are treated separately. This means that the first response time corresponds to the first change of the signal as seen from Figure 6. Another advantage is that the regression model allows for multiple response times, which could possibly represent different types of driving from the magnetosphere as discussed in section 6. By finding the response times which give the optimal fit between model and data, it is the goal that the signal should be enhanced and the noise reduced.

As with any statistical model, there is the possibility of overfitting. Using additional response times will always decrease the error term in equation (2). However, these response times may not represent physical processes.

We have tried to avoid this by setting strict criteria on the confidence intervals of the coefficients as described in section 5. However, we can not be completely sure that all the response times represent physical processes. This would in particular be true when there are other factors which influence the magnetic perturbations observed by AMPERE. For example, the total reconnection of magnetic flux on the dayside will be equal to the reconnection rate times the length of the reconnection X line. *Milan et al.* [2007] have found that the length of the X line can vary from event to event. If it varies on the same timescales as the reconnection rate, the coefficients in the regression model must compensate for this variation, and this would complicate the interpretation of the model.

In the rest of this section we will go through some important factors which are known to influence the observations from AMPERE.

### 7.1.1. IMF $B_Y$

By combining MLT sectors from each side of the noon-midnight meridian, we have implicitly assumed that the convection pattern is symmetric. However, it is well known [e.g., *Pettigrew et al.*, 2010] that the entire convection pattern is rotated for nonzero IMF $B_Y$. This rotation was explained to be due to the stress on newly opened flux tubes on the dayside by *Cowley et al.* [1991], and *Tenfjord et al.* [2015] recently showed how IMF $B_Y$ dynamically influences the entire magnetosphere in an MHD simulation. They also showed how the R1 and R2 current systems in AMPERE are rotated with about 1 MLT due to IMF $B_Y$. The results shown in this paper primarily involve a difference in the response times at the dayside and the nightside, and a rotation of 1 MLT will not significantly influence these results. However, it might be that the rotation is larger for strong $B_Y$. Therefore, we have repeated the analysis without intervals with strong IMF $B_Y$. It turned out that when we removed the strong $B_Y$ intervals, we also removed the intervals with the strongest reconnection rates. This resulted in longer average response times on the nightside. However, we were not able to identify any effects which could be attributed to IMF $B_Y$ only and have not shown these results here.

### 7.1.2. Other Convection Cells

There are some issues that have to be considered when using the direction of the magnetic field perturbations in the manner described in section 3. One is that this approach only considers the convection patterns for southward IMF. For northward IMF the patterns should be different [*Iijima*, 2000], with reversed perturbations poleward of the region 1 current system. Another issue is that there is a viscous convection cell which would give opposite magnetic field perturbations to the main convection cell [*Burch et al.*, 1985]. However, since the two-cell convection pattern for southward IMF is the dominating pattern [e.g., *Ruohoniemi and Greenwald*, 1996, and references therein], we expect our method to work in a statistical sense. The clear signature of a two-cell pattern in Figure 1 justifies this. In addition, *Korth et al.* [2005] found that viscous interaction only have a small contribution to the total convection. To make sure that our results are not influenced by northward IMF, we redid the analysis without the events with strongly positive IMF $B_Z$. This did not significantly alter any of the results.

### 7.1.3. Substorms

Apart from dayside reconnection, the strength of ionospheric convection also depends upon nightside reconnection [*Cowley and Lockwood*, 1992]. *Grocott et al.* [2002] found a strong enhancement of the convection at the nightside after substorm onset. This agrees with the observations that most of the flux which is opened on the dayside is closed after substorm onset [e.g., *Shukhtina et al.*, 2005]. Although substorm onset is found to come at a delay of about 1 h after a southward turning in the solar wind, sudden southward turnings are not so common overall. Most often the reconnection rate varies gradually and in small steps. Since we fit the statistical model to all types of variations in the reconnection rate, substorm onset may occur at any delay after a change in the reconnection rate and complicate the response times.

One way to avoid this problem would be to exclude intervals with substorm activity as done by *Ridley et al.* [1998]. In our case that would remove too much data because substorms are so common. *Milan et al.* [2008] found the period between substorms to be about 3 h for a reconnection rate of 2 mV/m, and *Milan et al.* [2007] found nightside reconnection to be active for 36% of the time. Another approach could be to somehow estimate the nightside reconnection rate and include it in the model as another exogenous input parameter, for example, by the method of *Shukhtina et al.* [2009]. This would be a good idea for a future study.

### 7.1.4. Conductivity

It is well known that momentum and magnetic tension are transmitted by shear Alfvén waves from the magnetosphere to the ionosphere [e.g., *Song and Vasyliūnas*, 2014]. The Alfvén wave is reflected above the ionosphere, and the magnetic perturbation will be the sum of the incident and reflected wave. The resulting magnetic amplitude is proportional to the Pedersen conductance times the convection electric field [e.g., *Nishida*, 1964], which also could be obtained by integrating equation (1) with respect to height [*Vasyliūnas*, 2012]. The conductance is dependent on the solar zenith angle, the flux of solar EUV radiation, and particle precipitation [*Brekke and Moen*, 1993]. While the two first parameters vary smoothly in time, the conductance can vary by large amounts during tens of minutes due to variations in the particle precipitation. The convection velocity is mainly controlled by the dayside reconnection rate [*Siscoe and Huang*, 1985] which in turn depends on solar wind parameters [*Reiff et al.*, 1981]. We acknowledge that this implies some uncertainty to the interpretation of magnetic perturbations as convection as long as the time development of the ionospheric conductances are unknown.

## 7.2. Response and Reconfiguration

*Cowley and Lockwood* [1992] (CL) explained the connection between the convection in the lobes and in the polar cap. When a newly opened flux tube is added to the lobes, the normal stresses will push field lines into the lobe near the perturbed region and outward in the surrounding region. The newly opened flux tube corresponds to a bulging of the polar cap near noon. Conceptually, ionospheric convection is excited when the perturbed polar cap boundary (PCB) evolves into a circular equilibrium shape. This concept was based on the analytical model by *Siscoe and Huang* [1985], where they showed that the entire ionospheric convection pattern can be derived from the motion of the PCB. The same concept can be used to describe the gradual reconfiguration of the convection pattern. *Cowley and Lockwood* [1997] suggested that the perturbation to the equilibrium boundary will propagate from noon to midnight in 15–20 min. They explained this delay to be due to the time it takes for an open flux tube to move from the dayside reconnection site into the near-Earth tail. A numerical model of this process was given in *Lockwood and Morley* [2004]. In their implementation it is the distance between the equilibrium boundary and the PCB which determines the convection velocity at the PCB. Furthermore, it is the distribution of this distance with respect to local time which determines the convection pattern. *Morley and Lockwood* [2006] have also shown that a global onset of convection can be explained by the CL paradigm in certain cases. If there is already ionospheric convection from previous dayside reconnection, a new reconnection pulse will create a convection response which is seen almost simultaneously in MLT sectors from noon to midnight.

In this paper we have contributed to the physical argument of the CL paradigm by explaining the response and reconfiguration of the convection in the lobe and polar cap in terms of MHD waves in a cold plasma [*Tamao*, 1964]. The most important result of this analysis is that these processes can be better explained by variations in the magnetopause flaring angle, than by the tailward motion of the open flux tubes. Since the orientation of IMF largely controls where open flux is added to the lobes [*Cowley et al.*, 1991], the flaring will first start to increase near GSM $Y = 0$ for purely southward IMF. During the next minutes the flaring spreads out toward the flanks of the magnetopause. This has some implications for the time development of the equilibrium boundary in the CL paradigm:

1. The delay between the responses near midnight and near noon is about 1 min corresponding to the travel time of a compressive wave through the lobe. This implies that the perturbation to the equilibrium boundary would propagate from noon to midnight on the same timescale.
2. The initial response is lower at the nightside because the amplitude of the compressive wave is reduced when it propagates through the lobe. This implies that the initial distance between the equilibrium boundary and the PCB would be smaller near midnight compared to near noon.
3. The reconfiguration of the convection pattern is due to redistribution of the magnetopause flaring toward the flanks of the magnetopause. The results reported here and in several studies in the past, suggest that this takes 10–20 min. During this time the distance between the equilibrium boundary and the PCB becomes more evenly distributed with respect to local times.
4. As the magnetopause flaring continues to increase, the magnitude of the convection also increases. This corresponds to a larger bulging of the polar cap near noon and an equatorward propagation of the equilibrium boundary.

Such an explanation is supported by the global MHD simulation by *Slinker et al.* [2001]. They found that compressive waves are responsible for a rapid convection response both on closed and open field lines. New in our interpretation is that we also consider the role of the evolution of the magnetopause flaring distribution. Our explanation does not contradict the explanation of *Morley and Lockwood* [2006], and both mechanisms may operate simultaneously contributing to rapid global onset of convection.

*Ridley et al.* [1998] suggested that a compressive wave in the ionosphere can explain the global response of convection. Since the ionosphere is almost incompressible, the fast propagation of the wave could give a response delayed by only a few seconds across the polar cap. However, as pointed out by *Lockwood and Cowley* [1999], while incompressibility implies that flow streamlines are closed loops, it does not imply that the same streamline connect both the dayside and the nightside. In the numerical examples by *Lockwood and Morley* [2004] it is shown that the convection can perfectly well be localized to the dayside. A question has also been raised by *Strangeway and Raeder* [2001] whether the horizontal energy transport in the ionosphere can be large enough to drive a response on the nightside from the dayside. This is because an almost incompressible ionosphere can support very little horizontal Poynting flux. This is not an issue in our interpretation since the Poynting flux is coming from the magnetosphere.

The model used in this paper is similar to the one derived by *Goertz et al.* [1993] for the ionospheric electric field. By considering the propagation of shear Alfvén waves along the geomagnetic field lines, they found that the time evolution of the electric field is described as a first-order difference equation. They also determined the rise time giving the best fit based on the *AU* index, and found it to be 70 min. This is higher than the values found in this study. However, if we restrict our model to only the zero lag input reconnection rate, we also get rise times close to 70 min except in the three MLT sectors closest to noon.

## 8. Conclusions

1. There is an almost simultaneous magnetic field response to dayside reconnection in all MLT sectors. This can be explained by compressive MHD waves propagating very quickly through the magnetosphere. These waves couple with shear Alfvén waves which transport the magnetic perturbation to the ionosphere.
2. The initial response is strongest near noon, while the responses on the nightside are most significantly enhanced after about 10 min. This indicates a reconfiguration of the ionospheric convection pattern on a similar timescale. We suggest that this can be explained by a magnetospheric flaring distribution which gradually spreads out from a small region on the magnetopause.
3. The magnetic responses rise in 30–50 min. We suggest that the rise time is partially due to the reflection of shear Alfvén in the ionosphere and partially due to the gradual increase of the magnetopause flaring angle.
4. This work fits with, and expands upon, the Cowley-Lockwood paradigm [*Cowley and Lockwood*, 1992, 1997; *Lockwood and Morley*, 2004; *Morley and Lockwood*, 2006].

## Appendix A: Regression Models

Coefficients and lags for the regression models are given here. Table A1 gives the parameters for the model in the case study in section 4, while Tables A2 and A3 gives the parameters for the models in the statistical study in section 5.

**Table A1.** Regression Models for the Magnetic Responses With Three Lags of the Reconnection Rate for the Case Study[a]

| MLT Sectors | $c$ | $2l_1$[b] | $\lambda_1$[c] | $2l_2$[b] | $\lambda_2$[c] | $2l_3$[b] | $\lambda_3$[c] | $t_r$[b] | $B_A(\infty)$[d] |
|---|---|---|---|---|---|---|---|---|---|
| 10&14 | 0.948 | 6 | 2.9 | 56 | 1.9 | 26 | -2.1 | 37 | 52 |
| 09&15 | 0.934 | 6 | 6.2 | 4 | -3.4 | 82 | 1.3 | 29 | 63 |
| 08&16 | 0.928 | 8 | 2.1 | 78 | 1.8 | 38 | 1.3 | 27 | 73 |
| 07&17 | 0.954 | 6 | 5.4 | 0 | -3.0 | 88 | 1.7 | 42 | 88 |
| 06&18 | 0.875 | 10 | 4.9 | 88 | 3.6 | 38 | 2.2 | 15 | 85 |
| 05&19 | 0.931 | 14 | 2.3 | 86 | 1.8 | 26 | 1.8 | 28 | 85 |
| 04&20 | 0.948 | 14 | 3.3 | 76 | 3.8 | 74 | -2.4 | 37 | 90 |
| 03&21 | 0.903 | 14 | 3.5 | 58 | 3.1 | 76 | 1.54 | 20 | 83 |
| 02&22 | 0.907 | 8 | 2.2 | 58 | 3.1 | 22 | 1.4 | 20 | 72 |

[a]The MLT-sectors are given in the first column. Columns 2–8 give the coefficients ($c$ and $\lambda_i$) and the responses times ($2\lambda_i$) for equation (3). The ninth column gives the rise time ($t_r$) from equation (8), and the last column gives the asymptotic value of the magnetic field from equation (7) for a reconnection rate of $R = 1$ mWb/(m s).
[b]Units are in minutes.
[c]Units are in nT/(mWb/(m s)).
[d]Units are in nT.

**Table A2.** Regression Models for the Northern Hemisphere for the Statistical Study in the Same Format as Table A1

| MLT Sectors | $c$ | $2l_1$[a] | $\lambda_1$[b] | $2l_2$[a] | $\lambda_2$[b] | $2l_3$[a] | $\lambda_3$[b] | $t_r$[a] | $B_A(\infty)$[c] |
|---|---|---|---|---|---|---|---|---|---|
| 10&14 | 0.950 | 0 | 2.1 | 78 | 0.8 | —— | —— | 39 | 59 |
| 09&15 | 0.957 | 0 | 1.4 | 4 | 1.5 | —— | —— | 46 | 69 |
| 08&16 | 0.953 | 0 | 1.4 | 4 | 1.7 | 86 | 0.7 | 42 | 81 |
| 07&17 | 0.959 | 0 | 1.3 | 6 | 1.3 | 12 | 1.0 | 47 | 87 |
| 06&18 | 0.960 | 6 | 1.1 | 12 | 1.4 | 0 | 1.2 | 49 | 93 |
| 05&19 | 0.951 | 0 | 1.7 | 12 | 1.9 | 82 | 0.8 | 40 | 89 |
| 04&20 | 0.949 | 10 | 1.9 | 0 | 1.1 | 68 | 1.1 | 39 | 81 |
| 03&21 | 0.952 | 10 | 2.4 | 60 | 1.1 | —— | —— | 40 | 74 |
| 02&22 | 0.943 | 10 | 1.4 | 50 | 1.3 | 0 | 1.0 | 34 | 65 |

[a]Units are in minutes.
[b]Units are in nT/(mWb/(m s)).
[c]Units are in nT.

**Table A3.** Regression Models for the Southern Hemisphere for the Statistical Study in the Same Format as Table A1

| MLT Sectors | $c$ | $2l_1$[a] | $\lambda_1$[b] | $2l_2$[a] | $\lambda_2$[b] | $2l_3$[a] | $\lambda_3$[b] | $2l_4$[a] | $\lambda_4$[b] | $t_r$[a] | $B_A(\infty)$[c] |
|---|---|---|---|---|---|---|---|---|---|---|---|
| 10&14 | 0.931 | 0 | 1.3 | 42 | 0.7 | 4 | 1.0 | 74 | 0.7 | 28 | 55 |
| 09&15 | 0.939 | 0 | 2.1 | 10 | 1.0 | 64 | 1.0 | —— | —— | 32 | 67 |
| 08&16 | 0.948 | 0 | 1.0 | 4 | 2.3 | 76 | 0.8 | —— | —— | 37 | 78 |
| 07&17 | 0.953 | 2 | 1.4 | 8 | 2.0 | 70 | 0.6 | —— | —— | 41 | 84 |
| 06&18 | 0.954 | 2 | 1.5 | 8 | 2.1 | 54 | 0.6 | —— | —— | 43 | 90 |
| 05&19 | 0.954 | 8 | 2.3 | 0 | 0.9 | 54 | 1.0 | —— | —— | 43 | 87 |
| 04&20 | 0.950 | 8 | 1.8 | 18 | 0.8 | 74 | 0.7 | 2 | 0.6 | 39 | 78 |
| 03&21 | 0.956 | 6 | 2.0 | 46 | 1.1 | —— | —— | —— | —— | 45 | 71 |
| 02&22 | 0.954 | 10 | 1.0 | 24 | 1.2 | 0 | 0.8 | —— | —— | 43 | 64 |

[a]Units are in minutes.
[b]Units are in nT/(mWb/(m s)).
[c]Units are in nT.

**Acknowledgments**

The authors are thankful for inspirational discussions with J.W. Gjerloev and R. Strangeway. This study was funded by the Research Council of Norway under contracts 216872/F50 and 223252/F50 (CoE). We thank the AMPERE team and the AMPERE Science Center for providing the Iridium-derived data products. The AMPERE data are available at ampere.jhuapl.edu. We acknowledge the Cluster FGM PI Chris Carr, the Cluster CIS PI Iannis Dandouras, and ESA Cluster Science Archive for making the Cluster data available.